\documentclass[aps,twocolumn,showpacs,pra,superscriptaddress,amsmath,amssymb]{revtex4-1}
\ifx\pdfoutput\undefined
\usepackage[dvipdfmx]{graphicx}
\usepackage[dvipdfmx]{hyperref}
\else
\usepackage{graphicx}
\usepackage{hyperref}
\fi

\usepackage{dcolumn}
\usepackage{bm}
\usepackage{braket}
\usepackage[usenames]{color}

\usepackage[varg]{txfonts}
\usepackage[T1]{fontenc}
\usepackage{xspace}

\begin{document}
\title{
Phase transitions in the Hubbard model for the bismuth nickelate
}

\author{Shoya Kojima}
\affiliation{
  Department of Physics, Tokyo Institute of Technology,
  Meguro, Tokyo 152-8551, Japan
}
\author{Joji Nasu}
\affiliation{
  Department of Physics, Tokyo Institute of Technology,
  Meguro, Tokyo 152-8551, Japan
}
\author{Akihisa Koga}
\affiliation{
  Department of Physics, Tokyo Institute of Technology,
  Meguro, Tokyo 152-8551, Japan
}

\date{\today}

\begin{abstract}
We study low temperature properties of the Hubbard model
for the bismuth nickelate, where degenerate orbitals in the nickel ions
and a single orbital in the bismuth ions are taken into account,
combining dynamical mean-field theory with the continuous-time
quantum Monte Carlo method.
We discuss the effect of the attractive interactions 
to mimic the valence skipping phenomenon in the bismuth ions.
We demonstrate how the charge and magnetically 
ordered states are stable against thermal fluctuations.
It is furthermore clarified that
the ferromagnetically ordered and orbital ordered states 
are stabilized due to the presence of the orbital degeneracy 
at low temperatures.
The crossover between metallic and insulating states is also discussed.
\end{abstract}


\maketitle


\section{Introduction}
Transition-metal oxides provide the typical playground
for strongly correlated electron systems.
Interesting examples 
are the manganite $\rm La_{1-x}Sr_xMnO_3$~\cite{Tokura} and 
the ruthenate $\rm Sr_2RuO_4$~\cite{Maeno}, 
where remarkable phenomena have been observed 
such as the colossal magnetoregistance and triplet superconductivity.
These stimulate extensive theoretical and experimental
investigations~\cite{Imada,Rice}.
Common physics in the perovskite-type oxides is that
the orbital degrees of freedom
in the transition-metal ions (Mn, Ru) play a crucial role for realizing
interesting low temperature properties.
On the other hand,
the lanthanum or strontium ions simply control electron number and
play little role for interesting phenomena.

Recently, the bismuth nickelate $\mathrm{BiNiO_3}$ has attracted
current interest since the successful observation of
the colossal negative thermal expansion~\cite{Azuma}.
In the compound, the phenomenon occurs together with
the valence transition in the nickel and bismuth ions.
In fact, it has been reported that, as increasing temperature and/or pressure,
the electronic configuration
$\mathrm{Bi^{3+}_{0.5}Bi^{5+}_{0.5}Ni^{2+}O^{2-}_3}$
is suddenly changed to that $\mathrm{Bi^{3+}Ni^{3+}O^{2-}_3}$.
One of the important points in the compound is the orbital degeneracy
in the nickel ions.
This may induce interesting low temperature states such as the orbital ordered
and ferromagnetically ordered states, which have not been observed up to now.
Another is that
the bismuth forms the compound with 3+ and 5+ valence states,
so called, a ``valence-skipping'' ion.
This behavior should be explained by considering
fairly large attractive interactions between $6s$ electrons~\cite{Varma}
although one naively expects that
they are little correlated,
comparing with $3d$ electrons in the nickel ions.
Therefore, it is necessary to treat electron correlations in both ions
correctly to understand low temperature properties in the compound.

Naka et al. have studied the single-band Hubbard model
to discuss low temperature properties in the compound
by means of the Hartree-Fock approximations~\cite{Naka}.
The existence of the valence transition has been clarified,
by considering the effective attractive interactions in the bismuth ions.
However, the orbital degeneracy in the nickel ions
has not been treated,
and thereby low temperature properties in the compound still remain unclear.
In particular,
the orbital degrees of freedom and the Hund coupling
should play a crucial role
to understand the valence and magnetic transitions in the system.
Therefore, it is necessary to treat the valence-skipping in the bismuth ions
as well as the orbital degeneracy in the nickel ions on equal footing.

To clarify the above issues,
we study the Hubbard model composed of two distinct sites with 
degenerate orbitals for nickel ions and
a single orbital for bismuth ions.
Here, using dynamical mean-field theory~\cite{DMFT1,DMFT2,DMFT3},
we discuss how local electron correlations in both ions affect
low temperature properties.
We then reveal the stability of the charge density wave (CDW) and
antiferromagnetically (AFM) ordered states
against thermal fluctuations.
Furthermore, we find the ferromagnetically (FM) ordered state and
the antiferroorbital (AFO) ordered state.
Examining low energy properties in the density of states, 
we discuss the metal-insulator crossover in the system. 

The paper is organized as follows.
In Sec.~\ref{sec:2}, we introduce
the model for the compound $\rm BiNiO_3$
and summarize our theoretical approach.
In Sec.~\ref{sec:att},
we discuss how the attractive interaction effectively realizes
the valence skip in the bismuth ions.
In Sec.~\ref{sec:3}, we obtain the phase diagrams to discuss
the role of the degenerate orbitals at low temperatures.
A summary is given in the last section.

\section{Model and Method}\label{sec:2}
\begin{figure}[t]
\begin{center}
\includegraphics[width=0.7\columnwidth]{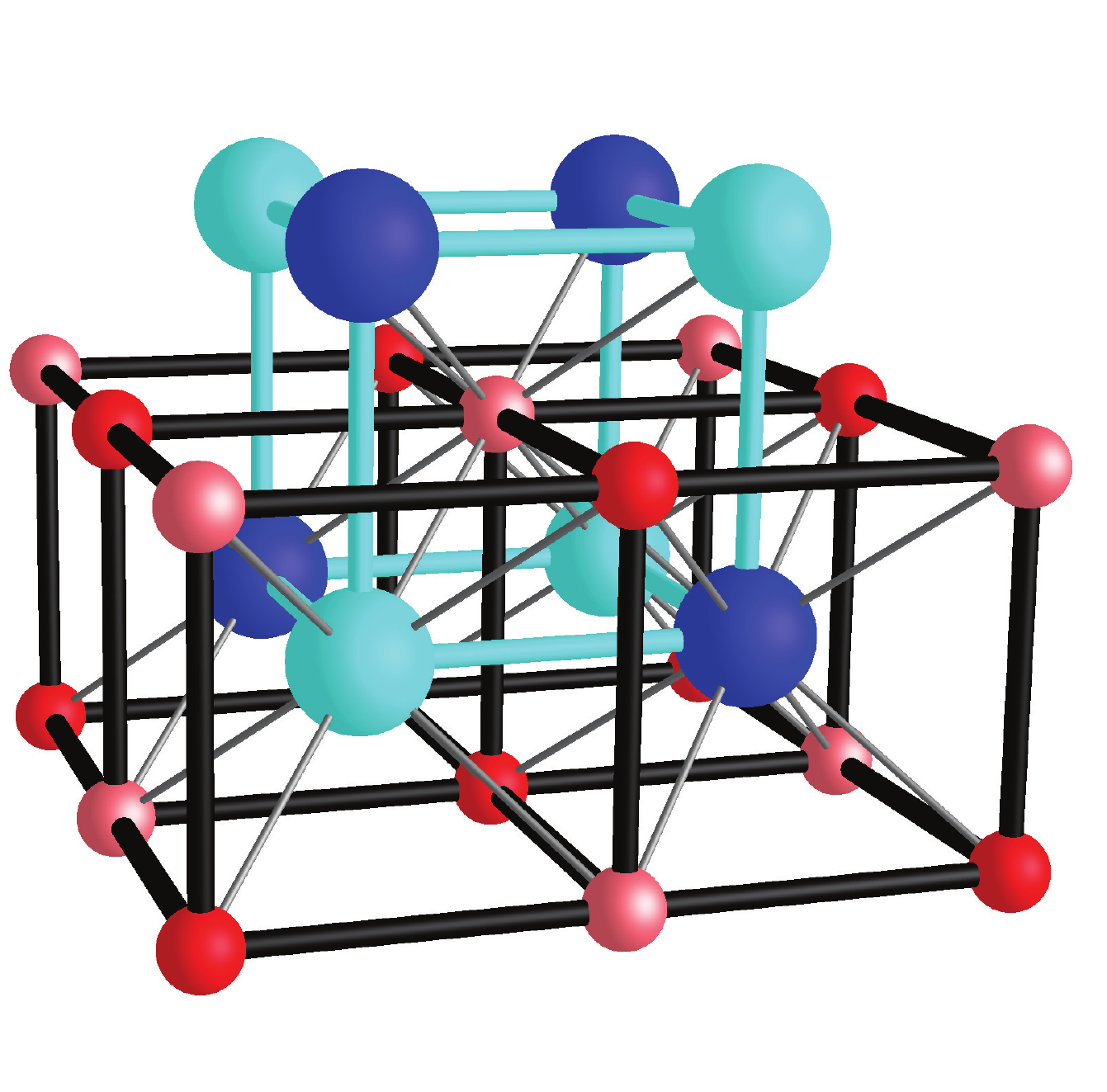}
\caption{
Three dimensional lattice structure in the compound $\rm BiNiO_3$.
Grayscaled large (small) spheres represent the Bi (Ni) ions
in the sublattice.
The hopping integrals $t_M$ ($t'$) between the same (different) ions
are expressed by the bold (thin) lines.
}
\label{fig:pic}
\end{center}
\end{figure}

We study the effect of strong electron correlations
in the compound $\rm BiNiO_3$.
The model should be described by the following Hubbard Hamiltonian,
\begin{align}
H & = H_{\rm Bi} + H_{\rm Ni} + H_{\rm Bi-Ni},\label{H}\\
H_{\rm Bi} &= -t_{\mathrm{Bi}}\sum_{\langle ij \rangle\sigma}^{\rm Bi}
\Bigl( b^{\dag}_{i \sigma} b_{j \sigma} +  \mathrm{H.c.} \Bigr)
+U_\mathrm{Bi}\sum_{i}^{\rm Bi}\left(n^\mathrm{Bi}_{i\uparrow}
-\frac{1}{2}\right) 
\left(n^\mathrm{Bi}_{i\downarrow}-\frac{1}{2}\right),\\
H_{\rm Ni} &= -t_{\mathrm{Ni}}\sum_{\langle ij \rangle\alpha\sigma}^{\rm Ni} 
\Bigl( a^{\dag}_{i \alpha \sigma} a_{j \alpha \sigma} +  \mathrm{H.c.} \Bigr)
  + \Delta\sum_{i\alpha\sigma}^{\rm Ni} n^\mathrm{Ni}_{i\alpha \sigma}\nonumber\\
&+ U_\mathrm{Ni} \sum_{i\alpha} \left(n^\mathrm{Ni}_{i\alpha \uparrow}-\frac{1}{2} \right)\left(n^\mathrm{Ni}_{i\alpha \downarrow}-\frac{1}{2}\right)\nonumber\\
  &+U'_\mathrm{Ni} \sum_{i\sigma} \left(n^\mathrm{Ni}_{i1\sigma}-\frac{1}{2}\right)\left( n^\mathrm{Ni}_{i2\bar{\sigma}}-\frac{1}{2}\right)\nonumber\\
 & +\Bigl( U'_\mathrm{Ni} - J \Bigr) \sum_{i\sigma}
\left(n^\mathrm{Ni}_{i1\sigma} -\frac{1}{2}\right)\left(n^\mathrm{Ni}_{i2\sigma}-\frac{1}{2}\right),\\
H_{\rm Bi-Ni}& = -t'\sum_{\langle ij \rangle \alpha \sigma}\Bigl( a^{\dag}_{i \alpha \sigma} b_{j \sigma} +  \mathrm{H.c.} \Bigr),
\end{align}
where
$b^\dagger_{i\sigma}$ ($b_{i\sigma}$)
is a creation (annihilation) operator of an electron with spin $\sigma$
at the Bi sites and
$a^\dagger_{i\alpha \sigma}$ ($a_{i\alpha \sigma}$)
is a creation (annihilation) operator of an electron
with spin $\sigma(=\uparrow, \downarrow)$ and
orbital index $\alpha(=1, 2)$ at the Ni sites of the $i$th unit cell,
$n^\mathrm{Bi}_{i\sigma}=b^\dagger_{i\sigma}b_{i\sigma}$, and
$n^\mathrm{Ni}_{i\alpha \sigma}=a^\dagger_{i\alpha \sigma}a_{i\alpha \sigma}$.
$t_M$ is the hopping integral
between $M(=$ Bi, Ni$)$ ions and $t'$ is
between Ni and Bi ions (see Fig.~\ref{fig:pic}).
$U_{\rm Bi}(<0)$ is the onsite interaction in the Bi ions, and
$U_{\rm Ni}$, $U'_{\rm Ni}$ and $J$ are the intraorbital interaction,
interorbital interaction, and Hund coupling in the Ni ions.
In our paper, we neglect the spin flip and pair hopping terms, for simplicity.
$\Delta$ is the energy difference
between an $e_g$ level in Ni ions and an $s$ level in Bi ions,
which should be tuned experimentally by the applied pressure.
The effective model may be regarded
to have the spatially alternating interactions~\cite{Saito}
since electron correlations are taken into account
in both nickel and bismuth ions.

The compound $\rm BiNiO_3$ exhibits the CDW and
AFM ordered states below the critical temperatures
$T_c^{\rm CDW}$ and $T_c^{\rm AFM}$,
where charges in Bi ions and spins in Ni ions are ordered,
respectively~\cite{Azuma}.
Therefore, the lattices for Ni and Bi ions are assumed 
to be divided into two sublattices,
which is schematically shown in Fig.~\ref{fig:pic}.
Each $M$ ion in a sublattice $A$ has nearest neighbor $M$ ions
in a sublattice $B$. 
On the other hand, each $M$ ion is surrounded by $\bar{M}$ ions
in both sublattices.
Therefore, the hopping $t'$ between different ions should
yield a sort of frustration,
which may suppress charge and magnetic fluctuations.

To discuss the stability of the ordered states,
we make use of DMFT~\cite{DMFT1,DMFT2,DMFT3}.
In DMFT, the original lattice model is mapped to an effective impurity model,
where local electron correlations are accurately taken into account.
The lattice Green's function is obtained via a self-consistency condition
imposed on the impurity problem.
This treatment is exact in infinite dimensions, and
the DMFT method has successfully explained
interesting physics such as
Mott transitions~\cite{Caffarel,Pruschke,Sakai,Bulla,Georges,Ono,Koga2band}
and magnetic transitions~\cite{Chitra,Zitzler,Momoi,Yanatori}.
When the realistic condition is concerned,
$t_M$ is more dominant than $t'$,
which allows us to use
the self-consistent equations for DMFT~\cite{Chitra,Zitzler} as
\begin{eqnarray}
\left({\mathcal{G}_{M\gamma}}\right)^{-1} =  i\omega_n +\mu_M
-\left(\frac{D_M}{2}\right)^2  G_{M\bar{\gamma}}
-\frac{1}{2}\left(\frac{D'}{2}\right)^2
\sum_{\gamma'} G_{\bar{M}\gamma'},
\label{eq:self}
\end{eqnarray}
where $\omega_n=(2n+1)\pi T$ is the Matsubara frequency,
$T$ is the temperature, $\mu_{\rm Ni}=\mu-\Delta$, $\mu_{\rm Bi}=\mu$, and
$\mu$ is the chemical potential.
${\mathcal{G}_{M\gamma}}$ ($G_{M\gamma}$) is
the noninteracting (full) Green's function for the $M$ ion
in the $\gamma(=A, B)$th sublattice.
For convenience, we neglect spin and orbital indexes in Eq. (\ref{eq:self}),
and consider the Bethe lattice
with coordination $z\rightarrow \infty$
as a lattice structure,
where the hoppings are rescaled as $D_M=2\sqrt{z}t_M$ and $D'=2\sqrt{z}t'$.
In the framework of DMFT, we solve effective impurity models
for Bi and Ni sites.
To discuss finite temperature properties in the system,
we employ the strong-coupling version of
the continuous-time quantum Monte Carlo method~\cite{Werner,solver_review},
which is efficient to study the Hubbard model in both weak and strong coupling
regimes.

There are some possible ordered states at low temperatures.
In the bismuth sites, the attractive interaction
$U_{\rm Bi}(<0)$ should induce the CDW state.
As for the nickel sites, the repulsive interactions
($U_{\rm Ni}, U_{\rm Ni}', J$) should
stabilize the ferromagnetically (FM) or AFM ordered
state.
Furthermore, the ferroorbital (FO) or antiferroorbital (AFO) ordered state
is also realizable, depending on the local electron fillings.
To characterize the above states, we here define the order parameters as,
\begin{align}
m^{\rm Bi}_{\rm CDW}&=\frac{1}{2}\langle n^{\rm Bi}_A-n^{\rm Bi}_B\rangle,\\
m^{\rm Ni}_{\rm FM(AFM)}&=\frac{1}{4}\langle S^z_A\pm S^z_B\rangle,\\
m^{\rm Ni}_{\rm FO(AFO)}&=\frac{1}{2}\langle T^z_A\pm T^z_B\rangle,
\end{align}
where $n^\mathrm{Bi}_\gamma = \sum n^{\rm Bi}_{\gamma \sigma}/2$ is
a local filling at the Bi sites,
$S^z_\gamma = \sum_\alpha \left(n^{\rm Ni}_{\alpha\uparrow\gamma}
- n^{\rm Ni}_{\alpha\downarrow\gamma}\right)/2$,
and
$T^z_\gamma = \sum_\sigma\left(n^{\mathrm{Ni}}_{1\sigma\gamma}
- n^{\mathrm{Ni}}_{2\sigma\gamma}\right)/2$
are the $z$-component of the spin and pseudospin
at Ni sites in the $\gamma$th sublattice.
Here, we focus on the above diagonal orders
to discuss their stabilities.
We also calculate the quantity as,
\begin{equation}
\tilde{A}_M=-\frac{1}{\pi T}G_M \left( \frac{1}{2T} \right).
\end{equation}
This quantity is reduced to the density of states for $M$ ions 
at the Fermi level in the limit $T\rightarrow 0$~\cite{SpectralWeight},
which should allow us to discuss how the metallic state 
$(\tilde{A}\neq 0)$ competes with the insulating state $(\tilde{A}=0)$
at finite temperatures.
In the paper, we use the hopping $D(=D_{\rm Ni}=D_{\rm Bi})$ 
as the unit of energy and 
fix the parameters as $U_\mathrm{Ni}=U'_\mathrm{Ni}+2J$, 
$U_\mathrm{Ni}/J=0.1$, and $D'/D=0.28$.
The total electron density is fixed as 
$\sum_\sigma (n_{\sigma\gamma}^{\rm Bi}
+\sum_\alpha n_{\alpha\sigma\gamma}^{\rm Ni})=2$
for each sublattice $\gamma$.
The model Hamiltonian Eq. (\ref{H}) is equivalent under the particle-hole
transformations except for the sign of the parameter $\Delta$.
Therefore, our discussions are restricted to the case
with $\Delta \ge 0$ without loss of generality.


\section{Effect of attractive interactions}\label{sec:att}
We discuss
the effect of the attractive interaction $U_{\rm Bi}$.
It is known that the bismuth ions behave as ``valence skipper''
in the compound~\cite{Azuma}.
To discuss how such a condition is realized by the attractive interaction,
we examine empty, singly, and doubly occupied states 
in the Bi site at certain parameters.
\begin{figure}[t]
\centering
\includegraphics[width=0.8\columnwidth]{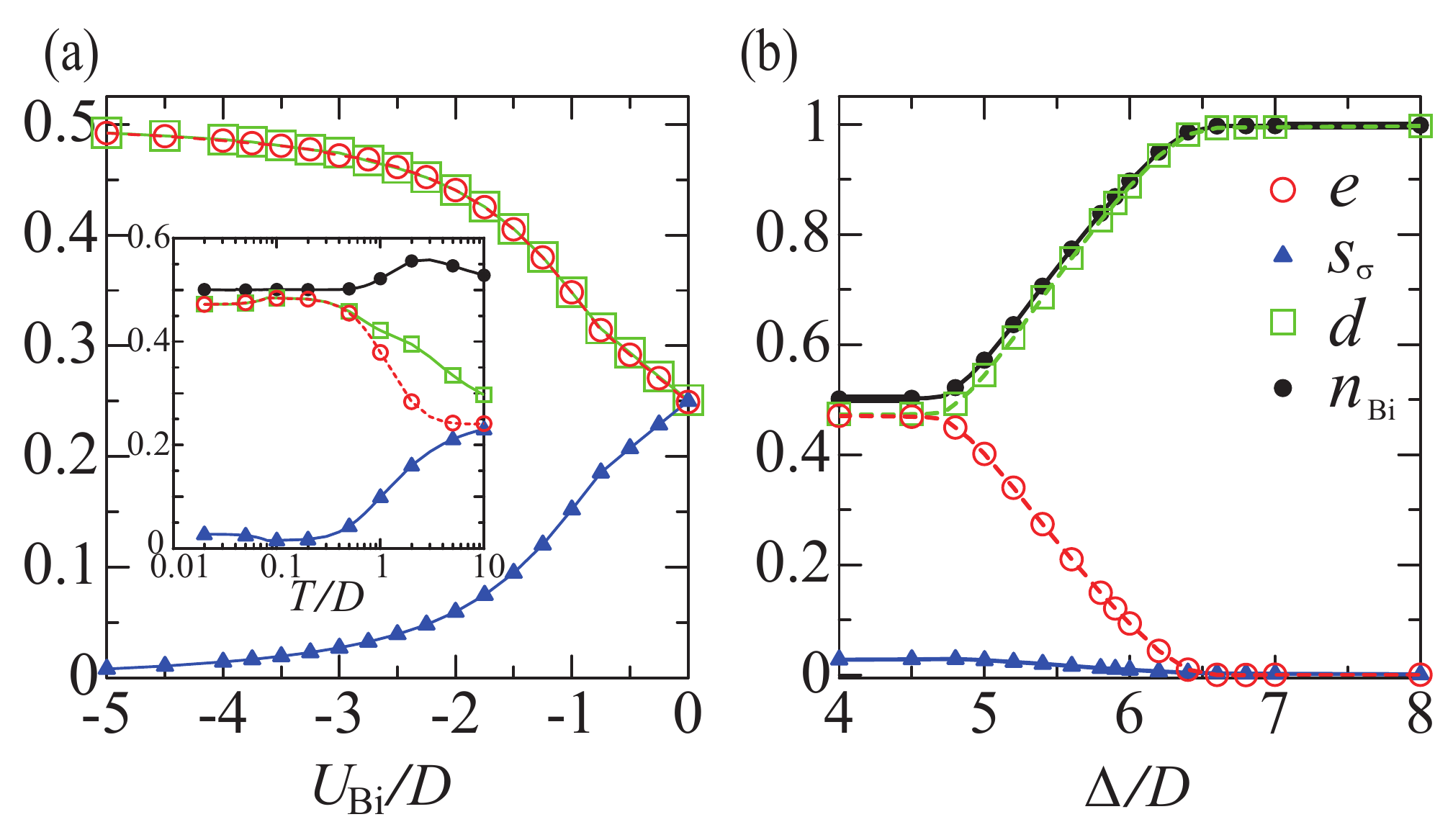}
\caption{
(a) The probabilities for empty, singly, and
doubly occupied states in the Bi states
in the system with $U_{\rm Ni}/D=10$ and $\Delta/D=2$
when $T/D=0.04$ (main panel) and $U_{\rm Bi}/D=-3$ (inset).
(b) The probabilities as a function of $\Delta$ when
$U_{\rm Ni}/D=10$, $U_{\rm Bu}/D=-3$, and $T/D=0.04$.
}
\label{fig:Occupancy}
\end{figure}
Figure \ref{fig:Occupancy}(a) shows their probabilities as a function
of the attractive interaction $U_{\rm Bi}$.
In the case,
the local band filling for Bi sites are always half filled (not shown).
When $U_{\rm Bi}=0$,
all possible states are equally realized with $e=s_\sigma=d=1/4$.
The introduction of the attractive interaction favors
empty and doubly occupied states, while decreases the probability
for singly occupied states.
It is found that the singly occupied state is little realized
($s_\sigma<0.04$) when $U_{\rm Bi}/D<-2.5$.
We show the temperature dependence in the system with $U_{\rm Bi}/D=-3$
in the inset of Fig.~\ref{fig:Occupancy}(a).
At high temperatures ($T>|U_{\rm Bi}|$),
the local band filling is away from half filling and
the single occupied states are realized with $s_\sigma\sim 0.2$.
Therefore, we can not discuss the valence skipping phenomenon in the case.
On the other hand, when $T/D<0.1(\ll|U_{\rm Bi}/D|)$,
the single occupancy is negligible and little depends on the temperature.
Furthermore, the single occupancy little depends 
on the energy difference $\Delta$ although
the local band filling for the Bi sites, and
the probabilities for the empty and doubly occupied states
are gradually changed, as shown in Fig.~\ref{fig:Occupancy}(b).
Therefore,
the Bi site can be regarded as the valence skipping one
when we fix the attractive interaction as $U_{\rm Bi}/D=-3$
and focus on low temperature properties $(T/D<0.1$).
The attractive interaction $U_{\rm Bi}/D=-1$ used
in the previous paper~\cite{Naka} should not be enough to capture
the valence skipping phenomenon in the Bi sites.

In the following, we fix the attractive interaction $U_{\rm Bi}/D=-3$.
Then, we discuss low temperature properties 
in the weak and strong coupling cases with
$U_{\rm Ni}/D=1$ and $U_{\rm Ni}/D=10$
where the experimental condition for the critical temperatures
$(T_c^{\rm AFM}< T_c^{\rm CDW})$ is satisfied.

\section{Low temperature properties}\label{sec:3}

\begin{figure}[t]
\centering
\includegraphics[width=0.8\columnwidth]{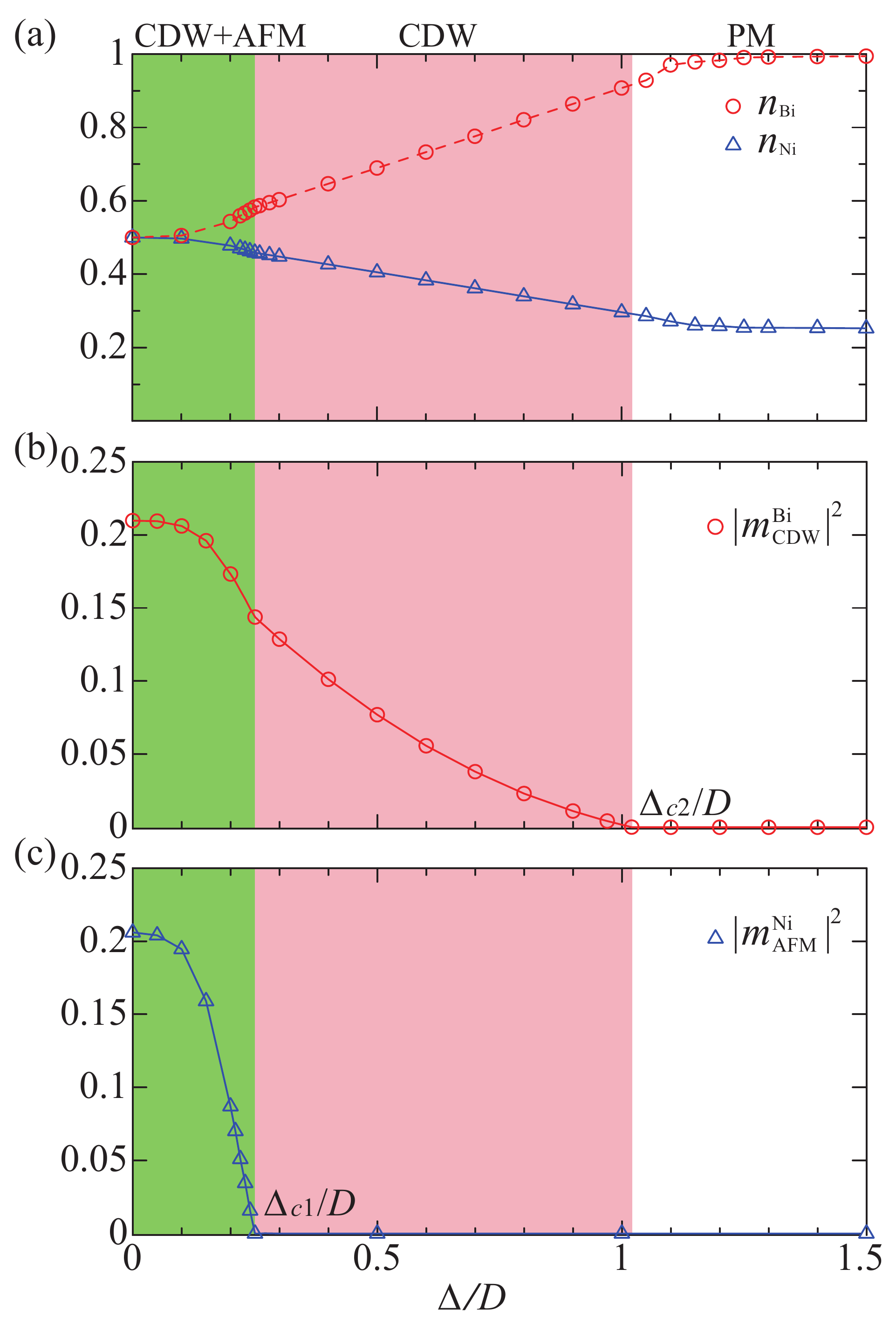}
\caption{
$\Delta$ dependences of (a) the local filling and the square of
the order parameters in (b) the Bi and (c) Ni sites as in the system with $U_{\rm Ni}/D=10$ when $T/D=0.02$.
}
\label{fig:DeltavsOrder_Weak}
\end{figure}

First, we would like to discuss low temperature properties
in the weak coupling case with $U_{\rm Ni}/D=1$.
By performing DMFT calculations at a temperature $T/D=0.02$,
we obtain the results, as shown in Fig.~\ref{fig:DeltavsOrder_Weak}.
When $\Delta=0$, both Ni and Bi ions are half filled
($n_{\rm Bi}=n_{\rm Ni}=0.5$).
We find that the CDW and AFM ordered states appear
with large order parameters $m^{\rm Bi}_{\rm CDW}=0.46$ 
and $m^{\rm Ni}_{\rm AFM}=0.45$.
Figures~\ref{fig:DeltavsOrder_Weak}(b) and~\ref{fig:DeltavsOrder_Weak}(c) show the square of the order
parameters, for discussions of the critical phenomena.
The magnetic moments in two orbitals of the Ni sites are parallel due to 
the existence of the Hund coupling.
Introducing $\Delta$, the number of electrons in Ni (Bi) sites
gradually decreases (increases), and both order parameters decrease.
At $\Delta=\Delta_{c1}$, the AFM order parameter vanishes and
the second-order phase transition occurs to a genuine CDW state.
We note that the hopping $t'$ simply suppresses 
magnetic and charge fluctuations, and
does not induce the cooperating phenomena between the Ni and Bi ions.
Therefore, the AFM and CDW states independently appear in the model
although a singularity appears in some curves of physical quantities.
Further increase in $\Delta$ decreases the CDW order parameter,
and finally $m_{\rm CDW}^{\rm Bi}$ vanishes at $\Delta=\Delta_{c2}$.
By examining critical behavior of the order parameters,
$m\sim |\Delta-\Delta_c|^\beta$ with the exponent $\beta=1/2$,
we obtain the critical values as $\Delta_{c1}/D\sim 0.25$ and 
$\Delta_{c2}/D\sim 1.0$, 
as shown in Figs.~\ref{fig:DeltavsOrder_Weak}(b) 
and~\ref{fig:DeltavsOrder_Weak}(c). 
In the large $\Delta$ region, 
the paramagnetic (PM) state is realized, 
where the Bi sites are almost fully occupied and 
the Ni sites are quarter-filled,
as shown in Fig.~\ref{fig:DeltavsOrder_Weak}(a).

\begin{figure}[t]
\centering
\includegraphics[width=0.8\columnwidth]{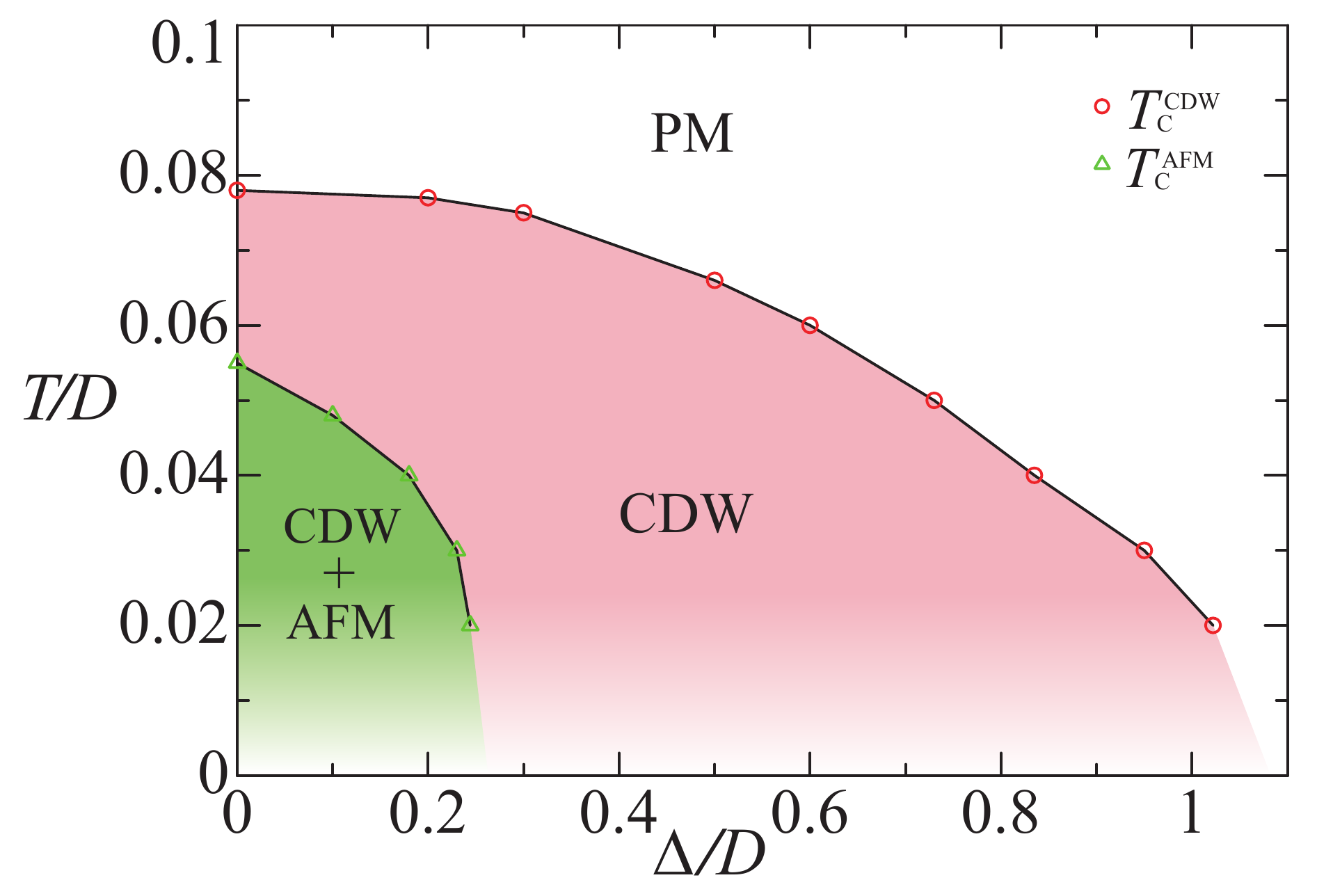}
\caption{
Phase diagram in the weak coupling case with
$U_\mathrm{Ni}/D=1$.
Circles (triangles) represent the second-order phase transition points,
where the CDW (AFM) order parameter vanishes.
}
\label{fig:PD_Weak}
\end{figure}

By performing similar calculations,
we obtain the phase diagram as shown in Fig.~\ref{fig:PD_Weak}.
We find that the CDW state is widely stabilized in the phase diagram,
where the empty and doubly occupied states are alternately realized
in the Bi sites.
On the other hand,
the AFM ordered state is realized in the Ni ions at lower $T$ and
smaller $\Delta$ region, as shown in Fig.~\ref{fig:PD_Weak}.
In the weak coupling case, the phase diagram is
essentially the same as that obtained
in the previous paper~\cite{Naka}.

\begin{figure}[t]
\centering
\includegraphics[width=0.8\columnwidth]{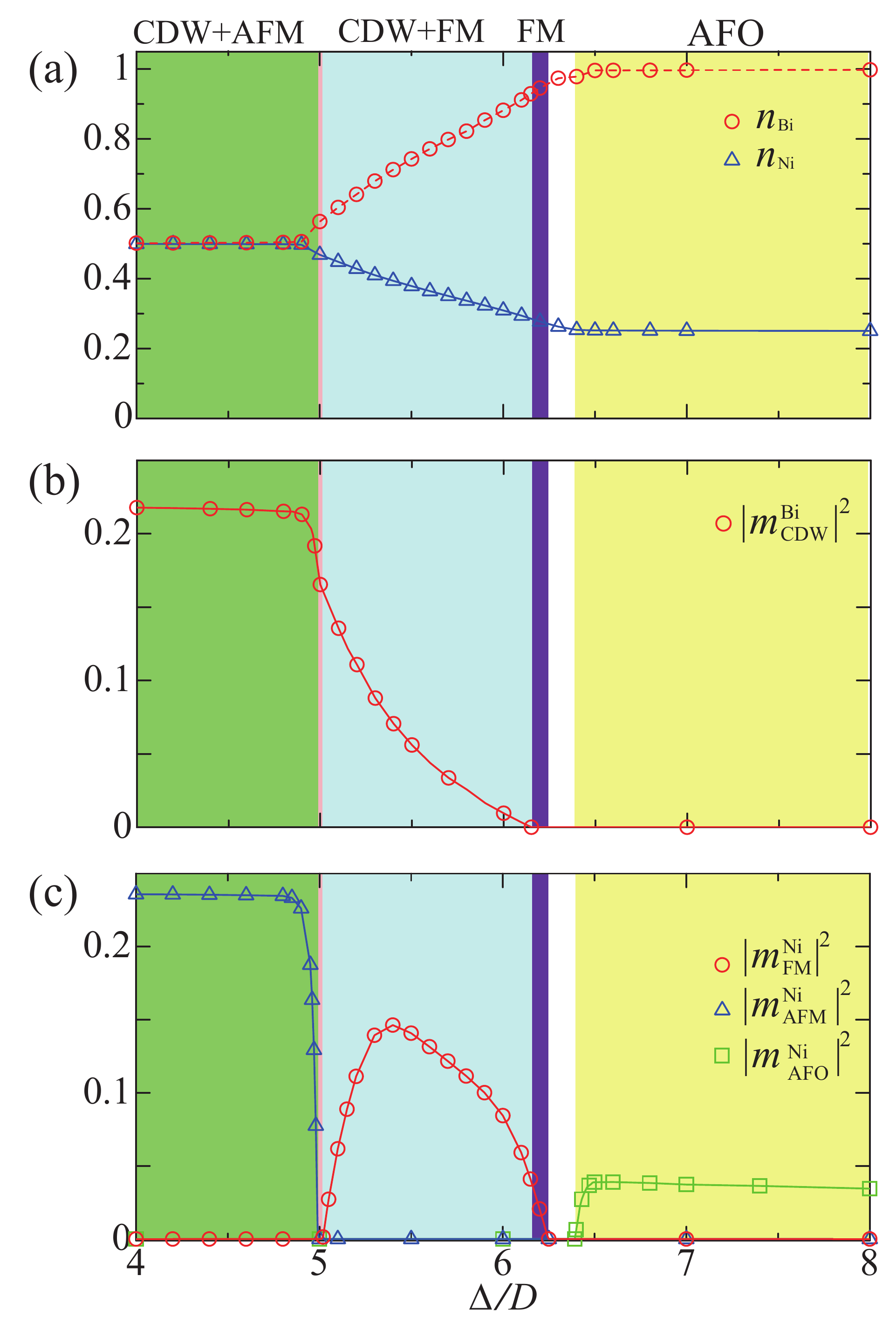}
\caption{
$\Delta$ dependences of (a) the local filling and the square of
the order parameters in (b) the Bi and (c) Ni sites in the system with $U_{\rm Ni}/D=10$ when $T/D=0.02$.
}
\label{fig:DeltavsOrder_Strong}
\end{figure}

Let us move to the strong coupling case with $U_{\rm Ni}/D=10$,
which may be relevant to the compound $\rm BiNiO_3$.
Figure~\ref{fig:DeltavsOrder_Strong} shows
the filling and order parameters at the low temperature $T/D=0.02$.
When $\Delta/D \lesssim 5$, both Bi and Ni sites are half-filled, 
and the CDW and AFM states are stabilized, as discussed above.
Around $\Delta/D\sim 5$, the local electron number is changed
and the staggered spin moment decreases rapidly.
This makes the AFM state unstable and the phase transition occurs 
in the Ni sites although the CDW state remains in the Bi sites.
At the temperature, this non-magnetic state is realized in a tiny region, 
which may be invisible in Fig.~\ref{fig:DeltavsOrder_Strong}.
Further increase in $\Delta$ induces the uniform spin moment
in the Ni sites $m^{\rm Ni}_{\rm FM}$, and the FM state is instead realized.
In the state, the number of electrons in the Ni site is intermediate and
no orbital moment appears. 
This is consistent with the fact that the ferromagnetic metallic ground state
is stabilized between quarter and half fillings 
in the degenerate Hubbard model~\cite{Momoi}.
When $n_{\rm Bi}\rightarrow 1$ and $n_{\rm Ni}\rightarrow 1/4$,
we find that the CDW and FM order parameters vanish 
at $\Delta/D=6.2$ and $\Delta/D=6.3$, respectively.
When $\Delta/D\sim 6.4$, the staggered orbital moment is induced rapidly and 
the AFO state is realized.
As the AFO state does not appear in the weak coupling case,
we conclude that the state is stabilized due to the strong Coulomb interaction
in the model with degenerate orbitals.
The larger value of $\Delta$ tends to fix the electron number as 
$n_{\rm Bi}=1$ and $n_{\rm Ni}=1/4$, and hence
the order parameter becomes constant.

\begin{figure}[t]
\centering
\includegraphics[width=0.8\columnwidth]{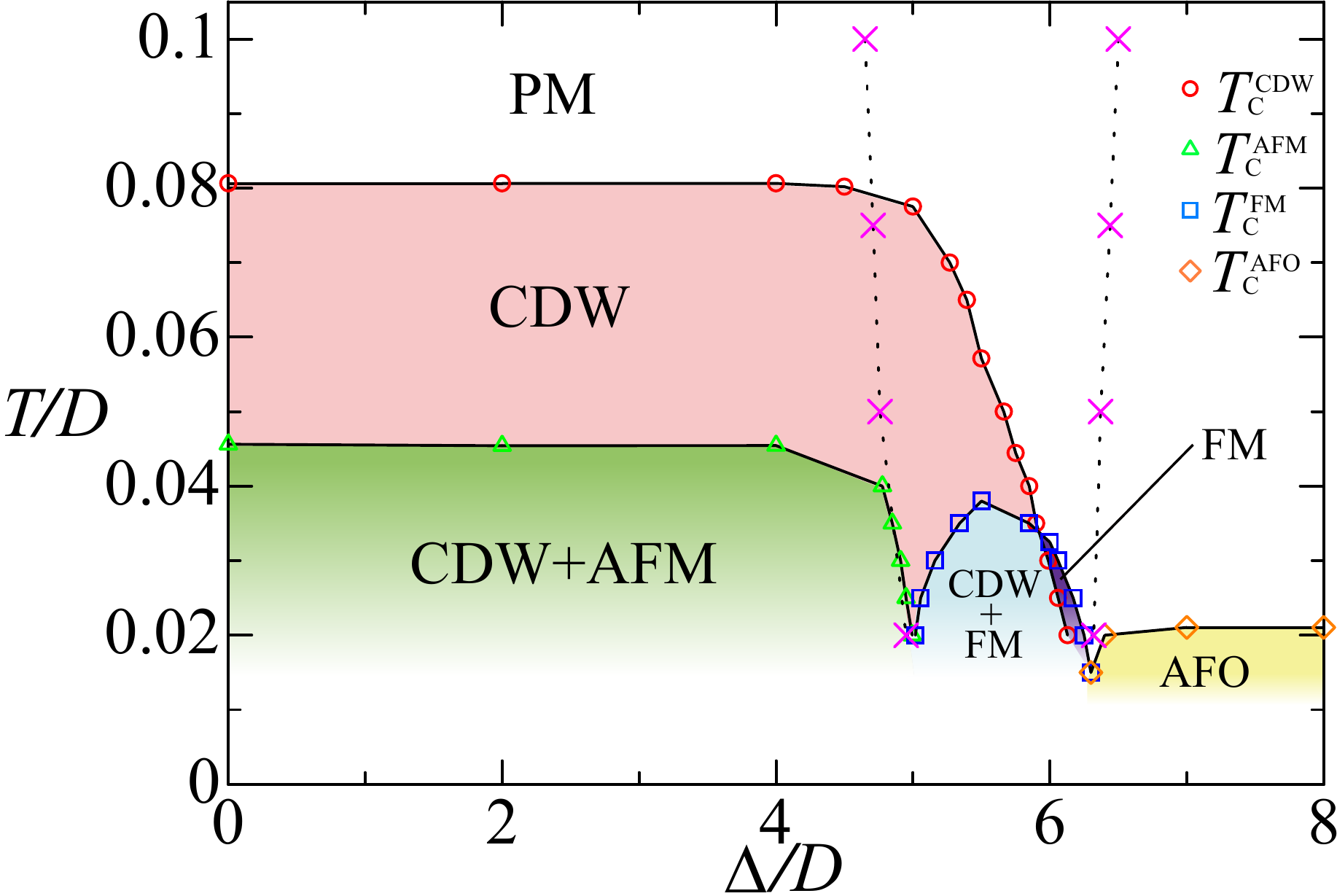}
\caption{
Phase diagram in the strong coupling case with
$U_\mathrm{Ni}/D=10$.
Circles, triangles, squares, and diamonds represent
the phase transition points,
where the CDW, AFM, FM, and AFO order parameter vanishes, respectively.
Crosses represent the metal-insulator crossover points (see text).
}
\label{fig:PD_Strong}
\end{figure}

The phase diagram is shown in Fig.~\ref{fig:PD_Strong}.
In the small $\Delta$ region, there appear the CDW and AFM states.
In the case, the critical temperatures are little changed
since the local band filling in each ion is fixed to be commensurate
due to strong electron correlations in Ni ions.
When $\Delta\sim 5.5 (\Delta > 6.5)$,
the FM (AFO) ordered states appears at low temperatures. 
It is expected that in the large $\Delta$ case,
the decrease of the temperature
enhances ferromagnetic correlations, which induces
the FM-AFO state with $m_{\rm FM}^{\rm Ni}\neq 0$ 
and $m_{\rm AFO}^{\rm Ni}\neq 0$~\cite{Momoi}. 
In the present study, we examine the Hubbard model with single-orbital sites
and degenerate-orbital sites, taking into account 
the valence skip and strong electron correlations.
The obtained phase diagram is consistent with the experiments~\cite{Azuma},
although the FM and AFO ordered states have not been observed.

\begin{figure}[t]
\centering
\includegraphics[width=0.8\columnwidth]{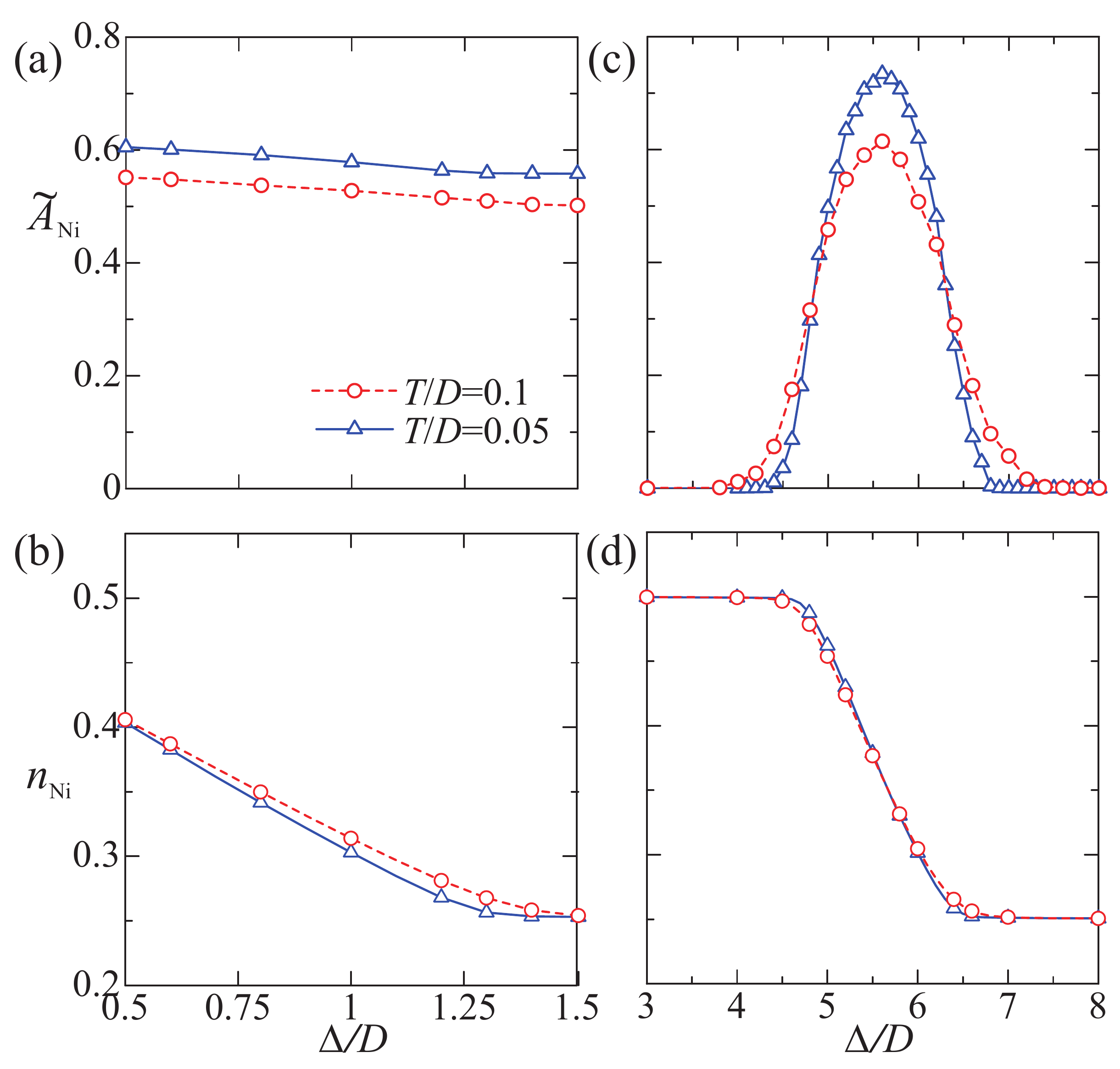}
\caption{
$\Delta$ dependences of (a) the density of states at the Fermi level $\tilde{A}_{\rm Ni}$
and (b) the local filling $n_{\rm Ni}$
at Ni sites in the weak coupling case with $U_\mathrm{Ni}/D=1$.
(c), (d) Corresponding data for the strong coupling case with $U_\mathrm{Ni}/D=10$.
}
\label{fig:DeltavsSpectralFunction}
\end{figure}

Finally, we discuss metallic state properties in the system.
When $T/D=0.05$,
the second-order phase transition occurs
between the CDW and PM states at $\Delta/D=0.73$ ($5.7$)
in the weak (strong) coupling case.
Here, we focus on metallic properties at the Ni sites in the system
close to the phase boundary for the CDW state.
The density of states at the Fermi level $\tilde{A}_{\rm Ni}$ and
the local band filling $n_{\rm Ni}$
are shown in Fig.~\ref{fig:DeltavsSpectralFunction}.
In the weak coupling case,
$\tilde{A}_{\rm Ni}$ is almost constant although
the local band filling $n_{\rm Ni}$ is gradually changed.
Therefore, the metallic state is always stable in the case.
On the other hand, in the strong coupling case,
different behavior appears.
When $\Delta/D\lesssim 4$, the local electron number at the Ni sites
is fixed as half filling.
As the Coulomb interaction at the nickel sites is fairly large,
we can say that
the Mott insulator is realized with $\tilde{A}_{\rm Ni}\sim 0$.
When $4.5\lesssim \Delta/D\lesssim 6.5$,
the intermediate filling is realized and metallic behavior appears
with finite $\tilde{A}_{\rm Ni}$.
Increasing $\Delta$, the local band filling approaches quarter
in the nickel sites. 
Then, Mott insulating behavior appears again 
with $\tilde{A}_{\rm Ni}\sim 0$, 
as shown in Fig.~\ref{fig:DeltavsSpectralFunction}(c).
Since we could not find any singularities 
in the curves, 
the crossover occurs between the metallic state and 
two Mott insulating states.
This crossover is not directly related to the phase transition in the Bi sites.
In fact, at higher temperatures $(T/D=0.01)$, 
the CDW state is not realized, but 
crossover behavior still remains,
as shown in Fig.~\ref{fig:DeltavsSpectralFunction}.
The crossover points, which may be deduced roughly under the conditions
$n_{\rm Ni}=0.26$ and $0.49$,
are shown as the crosses in Fig.~\ref{fig:PD_Strong}.
We find that the metal-insulator crossover little depends 
on the temperature, contrast to the phase boundary for the CDW state.
As decreasing temperatures, the crossover approaches the phase boundary 
for the FM state since the AFM and AFO states are realized 
only at the commensurate fillings and the FM state is realized 
in between, as discussed above.
Unfortunately, the valence transition
in the compound $\mathrm{BiNiO_3}$
between the electronic configurations
$\mathrm{Bi_{0.5}^{3+}Bi_{0.5}^{5+}Ni^{2+}O^{2-}_3}$ and
$\mathrm{Bi^{3+}Ni^{3+}O^{2-}_3}$ 
is of first order and the metal-insulator transition 
simultaneously occurs~\cite{Azuma,Ishiwata}.
This discrepancy may originate from the fact that
the valence transition in the compound is accompanied by
the structure phase transition associated with a volume shrinkage
or a strong mixing between the Ni $3d$ and ligand orbitals, 
which is often expected for perovskite nickelates~\cite{Ishikawa2005}.
Therefore,
the first-order phase transitions must be explained
when one considers the valence dependent hopping integrals, ligand orbitals,
electron-phonon coupling, etc,
which is now under consideration.

\section{Summary}
We have studied the Hubbard model for the compound $\rm BiNiO_3$,
where single (degenerate) orbital in the Bi (Ni) ions is taken into account.
Combining dynamical mean-field theory with the continuous-time
quantum Monte Carlo method,
we have discussed the finite temperature properties in the system.
It has been clarified how the valence skip phenomenon in the bismuth sites
is described by the attractive interactions.
By performing the DMFT calculations systematically,
we have obtained the phase diagrams, where
the ferromagnetically ordered and orbital ordered states appears
in addition to the charge density wave and antiferromagnetically
ordered states observed experimentally.
We have also discussed
the crossover between metallic and insulating states.

\begin{acknowledgments}
We would like to thank M. Naka for valuable discussions.
Parts of the numerical calculations are performed
in the supercomputing systems in ISSP, the University of Tokyo.
This work was partly supported by the Grant-in-Aid for
Scientific Research from JSPS, KAKENHI Grant Number 
25800193, 16H01066 (A.K.), and 16K17747 (J.N.).
The simulations have been performed using some of
the ALPS libraries~\cite{alps2}.
\end{acknowledgments}

\end{document}